\documentclass[letterpaper, 10 pt, conference]{ieeeconf}  % Comment this line out if you need a4paper

\IEEEoverridecommandlockouts                              % This command is only needed if 
\usepackage{graphics} % for pdf, bitmapped graphics files
\usepackage{amsmath} % assumes amsmath package installed
\usepackage{amssymb}  % assumes amsmath package installed

\newtheorem{theorem}{Theorem}

\newtheorem{remark}{Remark}

\newtheorem{definition}{Definition}
\renewcommand{\baselinestretch}{0.93}

\usepackage{cite}
\usepackage{subcaption}
\usepackage[nocomma]{optidef}
\usepackage[hyphens]{xurl} % ハイフン後改行優先
\makeatletter
\let\NAT@parse\undefined
\makeatother
\usepackage{hyperref}
\usepackage{color,comment}
\usepackage{hhline}
\usepackage{makecell,booktabs}
\usepackage{bm}
\title{\LARGE \bf
Koopman Temporal Logic and Model Predictive Control of\\
Nonlinear Systems: A Demonstration in AC-DC Power Conversion
}

\author{%
Shun Hirose, Shiu Mochiyama, and Yoshihiko Susuki
\thanks{This work was supported in part by JSPS KAKENHI (Grant No. 26K00968) and  JST BOOST (Grant No. JPMJBS2407). %and JSPS Bilateral Collaborations (Grant No. JPJSBP120242202). 
}
\thanks{They are with the Department of Electrical, Electronic, and Digital Science and Engineering, Kyoto University, Katsura, Nishikyo-ku, Kyoto 615-8510, Japan {\tt hirose.shun.65x@st.kyoto-u.ac.jp}, {\tt \{mochiyama.shiu, susuki.yoshihiko\}.5c@kyoto-u.ac.jp}}
}

\begin{document}

\maketitle
\thispagestyle{empty}
\pagestyle{empty}

%%%%%%%%%%%%%%%%%%%%%%%%%%%%%%%%%%%%%%%%%%%%%%%%%%%%%%%%%%%%%%%%%%%%%%%%%%%%%%%%
\begin{abstract}

This paper proposes a novel temporal logic, which we call Koopman Temporal Logic (KTL), towards a formal method in which Koopman operator theory provides a mathematical way to verify and control systems with nonlinear dynamics. 
We first review signal temporal logic and then introduce KTL, including its syntax, Boolean semantics, and quantitative semantics. 
Next, we propose a formulation of Model Predictive Control (MPC) from a control specification described by KTL. 
Effectiveness and computational efficiency of the proposed MPC formulation are evaluated by applying it to a reliable design of AC-DC power conversion.

\end{abstract}
\begin{keywords}
Koopman operator, Temporal logic, Model predictive control
\end{keywords}
\IEEEpeerreviewmaketitle

%%%%%%%%%%%%%%%%%%%%%%%%%%%%%%%%%%%%%%%%%%%%%%%%%%%%%%%%%%%%%%%%%%%%%%%%%%%%%%%%

\section{INTRODUCTION}

Formal synthesis, whose goal is to synthesize or control finite-state transition systems from a temporal logic specification, has attracted interests with connection to control of dynamical systems with real-valued states: see, e.g., \cite{FormalMethodsforControl,Lindemann-book}. 
Temporal logics have been widely utilized for specifying desired behaviors of transition systems. 
The so-called Signal Temporal Logic (STL) \cite{MonitoringTemporalProperties,RobustSatisfactionofTemporal} is a temporal logic for time-series signals and has been widely used in controller synthesis: see, e.g., \cite{raman2014,Lindemann2019}. 
The authors of \cite{raman2014} formulate a Model Predictive Control (MPC) with STL specifications for plants with linear dynamics. 
%{\color{red}The authors of \cite{penedo2018} extend STL for linear partial differential equations (PDEs), which they call \emph{Spatial-STL} (S-STL), and synthesize control from S-STL specifications using a finite element method.} 

The Koopman operator framework has recently been integrated with formal synthesis for controlling plants with nonlinear dynamics \cite{Miyashita2022,hirose2026formal}. 
The framework has received significant attention in nonlinear control theory and practice with a close connection to data-driven technologies: see, e.g., \cite{koopman-introduction-springer,BEVANDA2021,otto2021,Koopman-operators-in-robot-lerning,mezic2026koopmanoperatortheoryfundamentals}.
The so-called Koopman operator, which is a linear, infinite-dimensional representation of nonlinear systems \cite{operator-is-model}, is utilized to model and predict temporal behaviors of nonlinear plants. 
In \cite{Miyashita2022} we proposed to incorporate control specifications described by STL into the so-called {Koopman-MPC (K-MPC)} \cite{KORDA2018149} and demonstrated it for a heat management problem in a hospital. 
The K-MPC with STL specifications is demonstrated in other control problems in \cite{sekiguchi-kmpc-stl,hara-kmpc-stl}. 
In \cite{hirose2026formal} we extended the technique of \cite{Miyashita2022} for robust synthesis of the {K-MPC} by considering modeling error in the Koopman representation, which is inevitable through a finite-dimensional approximation, and demonstrated it for an AC-DC converter motivated by the More-Electric Aircraft (MEA) \cite{onboard}.

The objective of this paper is to reformulate the formal synthesis of nonlinear systems in \cite{Miyashita2022,hirose2026formal} by inventing a temporal logic suitable for dynamical system design. 
A practical motivation here is a computational issue arising in \cite{hirose2026formal}; The {K-MPC} with STL specifications poses an MIQP (Mixed-Integer Quadratic Programming) which is hard to solve, in particular, for large-scale problems. 
More fundamentally, to the best of the authors' survey, there is no temporal logic that directly verifies and monitors properties of nonlinear systems.  
For instance, STL can verify signals (or outputs) generated by a nonlinear system, but it is not direct to the underlying system itself. 
Note that the authors of \cite{penedo2018} extend STL for linear partial differential equations, which they call \emph{Spatial-STL} (S-STL), and synthesize control from S-STL specifications using a finite element method. 
In this paper, we build on the Koopman representation \cite{operator-is-model} to introduce a novel temporal logic for specifying properties of nonlinear systems, which we term as \emph{Koopman Temporal Logic} (KTL). 
We then utilize KTL to reformulate the MPC problem with STL specifications \cite{raman2014,Miyashita2022,hirose2026formal}.

\begin{comment}
This paper proposes a Koopman operator-based temporal logic, which specifies the temporal properties of the observables in a formal manner, and the control synthesis using the proposed temporal logic. 
As a case study, the proposed control synthesis is applied to an AC-DC converter. 
\end{comment}
The contributions of this paper are threefold.
\begin{itemize}
    \item[1)] A novel temporal logic is proposed to specify properties of systems with linear and nonlinear dynamics in Section~\ref{koopman-temporal-logic}. 
    The Koopman representation is roughly a pair $\{\mathcal{F}, \mathbf{K}\}$, where $\mathcal{F}$ is a function space, called observable space, and $\mathbf{K}$ is a linear operator acting on $\mathcal{F}$ \cite{operator-is-model}.
    We explicitly incorporate the pair $\{\mathcal{F},\mathbf{K}\}$ into the alphabet and predicates (see Section~\ref{koopman-temporal-logic}) of the proposed temporal logic. 
    Then, by combining predicates with STL-based grammar, we construct desired formulae to specify properties of an underlying nonlinear system. 
    As an important consequence, we show that the Koopman operator enables the system's verification, that is, to judge whether or not the nonlinear system satisfies a specification described by KTL. 

    \item[2)] An MPC formulation with KTL specifications is newly proposed in Section~\ref{MPC}. 
    Specifically, we leverage the Koopman operator to predict the value of quantitative satisfaction (see Section~\ref{koopman-temporal-logic}) of a KTL formula in the MPC framework. 
    We formulate an optimization problem subject to a finite-dimensional approximation of the Koopman representation that predicts not only the system's state but also the satisfaction value. 
    By comparison to the MPC with STL specifications \cite{raman2014}, the MPC formulation in this paper does not need any binary variable, implying that the resulting optimization problem is at most a QP (Quadratic Programming).

    \item[3)] The MPC formulation with KTL specifications is demonstrated in Section~\ref{application} for a reliable design of AC-DC power conversion. 
    We set KTL specifications for the AC current and DC voltage of an AC-DC converter and then formulate the optimization problem. 
    We show that the MPC formulation in this paper outperforms the conventional MPC with STL specifications in terms of computational cost, which can solve the computation issue in the initial motivation.
    
\end{itemize}
%
%The novelty of this paper is , to the best of authors' knowledge, the incorporation of formal methods into the Koopman operator theory.

\textit{Notation:} The symbol $\top$ stands for the transpose operation of vectors. 
The symbol $\mathrm{j}=\sqrt{-1}$ stands for the imaginary unit.
The symbols $\Im(\cdot)$ and $\Re(\cdot)$ stand for the imaginary and real part of $(\cdot)$. 
For a vector $v$, $v^{(i)}$ represents its $i$-th element.
The set of Boolean values is denoted by $\mathbb{B}:=\{\mathrm{true},\mathrm{false}\}$, and the set of discrete time indices by $\mathbb{T}=\mathbb{Z}_{\geq 0}:=\{0,1,2,\ldots\}$. 
The symbol $\mathcal{F}$ represents a Hilbert space. 
For $a, b \in \mathbb{Z}_{\geq 0}$ with $a \leq b$, $[a,b]$ stands for the discrete interval $\{a, a+1, \ldots, b\}$, and $x_{a:b}$ for the sequence $x_a, x_{a+1}, \ldots, x_b$. 

%%%%%%%%%%%%%%%%%%%%%%%%%%%%%%%%%%%%%%%%%%%%%%%%%%%%%%%%%%%%%%%%%%%%%%%%%%%%%%%%%%%%%%%%%%%%%%%%%%%%%%%%%%%%%%%%

\section{KOOPMAN TEMPORAL LOGIC} \label{koopman-temporal-logic}

This section presents our novel temporal logic, called the Koopman Temporal Logic (KTL).

\subsection{Review of Koopman Operator \cite{koopman-introduction-springer}}

Let us review the Koopman representation in terms of state-space formalism. 
Consider a discrete-time dynamical system given by 
\begin{equation}
    \label{eq:discrete-nonlinear-system}
    \bm{x}_{t+1}=\bm{T}(\bm{x}_{t}),
\end{equation}
where $\bm{x}_t\in\mathbb{R}^{n}$ represents the state at discrete time $t\in\mathbb{T}$ and $\bm{T}:\mathbb{R}^{n}\to\mathbb{R}^{n}$ represents a nonlinear map. 
As stated above, in the Koopman operator framework, we choose a scalar-valued function $f:\mathbb{R}^n\to\mathbb{R}$, called \emph{observable}, and analyze its time evolution, called the output $y$ of the nonlinear system \eqref{eq:discrete-nonlinear-system}:
\begin{equation*}
    y_t = f (\bm{x}_t).
\end{equation*}
The time evolution of an observable along the dynamical system \eqref{eq:discrete-nonlinear-system} is represented by the Koopman operator $\mathbf{K}$, defined as 
\begin{equation}
    \label{eq:koopman-definition}
    (\mathbf{K} f)(\bm{x}_{t})
    :=f\bigl(\bm{T}(\bm{x}_{t})\bigr)
    =f(\bm{x}_{t+1}).
\end{equation}
The operator $\mathbf{K}$ is linear in terms of its action onto the observable space $\mathcal{F}$. 
For a given observable $f_0\in\mathcal{F}$, we denote the time evolution of $f$ by $f_t:= f_0(\bm{T}^t(\bm{x}))$ and then derive the linear evolution $f_t$ as
\begin{equation*}
f_{t+1}=\mathbf{K}f_t.
%\label{eqn:Koopman_representation}
\end{equation*}
This is a foundation of the Koopman representation of the nonlinear system \eqref{eq:discrete-nonlinear-system}.
Under certain conditions, the Koopman operator $\mathbf{K}$ contains the full information about properties of the nonlinear system \eqref{eq:discrete-nonlinear-system}: see, e.g., \cite{koopman-introduction-springer}.

%%%%%%%%%%%%%%%%%%%%%%%%%%%%%%%%%%%%%%%%%%%%%%%%%%%%%%%%

\subsection{Review of Signal Temporal Logic \cite{MonitoringTemporalProperties,RobustSatisfactionofTemporal}}
 
Next, we review the Signal Temporal Logic (STL). 
STL is a temporal logic designed for a real-valued signal. 
A \emph{signal} $\bm{s}=\bm{s}_0\bm{s}_1\bm{s}_2\ldots$ is an infinite sequence, where $\bm{s}_t\in\mathbb{D}=\mathbb{R}^n$ is a real-valued vector at discrete time $t$.   
\begin{comment}
STL is a temporal logic defined over a time-series signal, called a \emph{trace}. 
Let the domain of the valuation of the signal $\bm{x}_t$ ($t=1,2,\ldots$) be $\mathbb{D}=\mathbb{R}^n$. 
%Let the state space be $\mathbb{D}=\mathbb{R}^n$. 
A trace is defined as a function $w$ mapping from $\mathbb{T}$ to $\mathbb{R}^n$, i.e., $w(t) = \bm{x}_t~(t=1,2,\ldots)$. 
%Here, we evaluate the state variable $\bm{x}$ of the dynamical system \eqref{eq:discrete-nonlinear-system}. 
\end{comment}
The grammar of STL is presented as follows.
\begin{definition}[Syntax] \label{def:stl-syntax}
    The basic formulae of STL are defined by the grammar 
    \begin{align*}
        \varphi \Coloneqq & \mu \mid \neg\,\varphi \mid \varphi_1 \land \varphi_2 \mid \varphi_1 \lor \varphi_2 \mid
        %\mid \mathbf{F}_{[a,b]}\varphi \mid \mathbf{G}_{[a,b]}\varphi \mid 
        \varphi_1 \,\mathbf{U}_{[a,b]}\, \varphi_2.
    \end{align*}
\end{definition}
In STL, an atomic proposition is interpreted as a \emph{predicate}, denoted by $\mu$. 
The predicate $\mu:\mathbb{D} \to \mathbb{B}$ is an interface between a signal and a truth value. 
The truth value of $\mu$ is determined by a sign of a function $\pi^\mu(\bm{s}_t)$, i.e., $\mu=\{\pi^\mu(\bm{s}_t)\geq0\}$, where $\pi^\mu:\mathbb{D}\to\mathbb{R}$ is a user-defined function. 
An STL formula is constructed by combining predicates with the Boolean and temporal operators. 
The notions $\neg$ (negation), $\land$ (conjunction), and $\lor$ (disjunction) are the standard Boolean operators. 
%The disjunction operator ($\lor$) can be derived from the negation and conjunction operators.
The notion $\mathbf{U}_{[a,b]}$ (\emph{until}) is a temporal operator. 
By using negation and the until operator, the two temporal operators,
\begin{equation*}
    \mathbf{F}_{[a,b]}\varphi = true\ \mathbf{U}_{[a,b]}\ \varphi, \quad \mathbf{G}_{[a,b]}\varphi = \neg \mathbf{F}_{[a,b]}\ \neg \varphi,
\end{equation*}
are derived. 
The notions $\mathbf{F}_{[a,b]}$ and $\mathbf{G}_{[a,b]}$ are called the \emph{eventually} operator and the \emph{always} operator. 

The satisfaction of an STL formula $\varphi$ by a signal $\bm{s}$ at discrete time $t$ is denoted by $(\bm{s},t)\models\varphi$. 
The following definition provides Boolean semantics for the STL's grammar in Definition~\ref{def:stl-syntax}. 
\begin{definition}[Boolean semantics] \label{def:stl-boolean-semantics}
    The satisfaction of an STL formula by a signal $\bm{s}$ at discrete time $t$ is inductively defined as follows:
    \begin{equation*}
        \begin{alignedat}{2}
        &(\bm{s},t)\models\mu
            &&\Leftrightarrow \pi^\mu(\bm{s}_t)\ge 0,\\
        &(\bm{s},t)\models\neg\varphi
            &&\Leftrightarrow (\bm{s},t)\not\models\varphi, \\
        &(\bm{s},t)\models\varphi_1\wedge\varphi_2
            &&\Leftrightarrow (\bm{s},t)\models\varphi_1\wedge(\bm{s},t)\models\varphi_2, \\
        &(\bm{s},t) \models \varphi_1 \mathbf{U}_{[a,b]} \varphi_2
            &&\Leftrightarrow \exists t'\in[t+a,t+b], (\bm{s},t') \models\varphi_2 \\
        & &&\hphantom{{}\Leftrightarrow{}}\wedge\forall t''\in[t,t'], (\bm{s},t'')\models\varphi_1. \\
        \end{alignedat}
    \end{equation*}
\end{definition}

The validity of an STL formula is quantitatively evaluated with a real-valued function, called a \emph{robustness function}. 
The STL formula $\varphi$ is satisfied by a signal $\bm{s}$ at discrete time $t$, i.e., $(\bm{s},t) \models \varphi$ if and only if the corresponding robustness function $\rho^\varphi$ is non-negative: $\rho^\varphi(\bm{s},t) \geq 0$. 
The robustness functions are constructed recursively in the following manner.
\begin{definition}[Quantitative semantics]
    \label{def:stl-quantitative-semantics}
    The complete robustness semantics for STL are defined as follows:
    \begin{equation*}
    \begin{aligned}
        \rho^{\mu}(\bm{s},t) &= \pi^\mu(\bm{s}_t),\\
        \rho^{\neg\varphi}(\bm{s},t) &= -\rho^{\varphi}(\bm{s},t),\\
        \rho^{\varphi_1\land\varphi_2}(\bm{s},t)
            &= \min\bigl(\rho^{\varphi_1}(\bm{s},t),\rho^{\varphi_2}(\bm{s},t)\bigr),\\
        \rho^{\varphi_1\mathbf{U}_{[a,b]}\varphi_2}(\bm{s},t)
            &= \max_{t'\in[t+a,t+b]}\min\Bigl(\rho^{\varphi_2}(\bm{s},t'),\\
            &\qquad
               \min_{t''\in[t,t']} \rho^{\varphi_1}(\bm{s},t'')\Bigr).\\
        % \rho^{\mathbf{F}_{[a,b]}\varphi}(w,k)
        %     &= \max_{k'\in[k+a,k+b]}\rho^{\varphi}(w,k'),\\
        % \rho^{\mathbf{G}_{[a,b]}\varphi}(w,k)
        %     &= \min_{k'\in[k+a,k+b]}\rho^{\varphi}(w,k').
    \end{aligned}
    \end{equation*}
\end{definition}
The value of the robustness function involving the temporal operator $\mathbf{U}_{[a,b]}$ at time $t$ depends on future values of the signal $\bm{s}$. 
The horizon length, defined recursively below, is the number of time steps required to compute the value of a robustness function at time $t$: $h^{\mu}=0$, $h^{\neg\varphi}=h^{\varphi}$, 
$h^{\varphi_1\wedge\varphi_2}=\max(h^{\varphi_1},h^{\varphi_2})$, and $h^{\varphi_1\mathbf{U}_{[a,b]}\varphi_2}=b+\max(h^{\varphi_1},h^{\varphi_2})$: see \cite{horizon-length-introduce} for details.  

%%%%%%%%%%%%%%%%%%%%%%%%%%%%%%%%%%%%%%%%%%%%%%%%%%%%%%%%

\subsection{Description of Koopman Temporal Logic}

We describe KTL, a novel temporal logic that combines STL with the Koopman operator. 
As reviewed above, the Koopman representation describes the time evolution of observables, $f_t=\mathbf{K}^tf_0$, where $\mathbf{K}$ contains the full information of the nonlinear system \eqref{eq:discrete-nonlinear-system}. 
As a natural extension, an \emph{observable}-valued signal $f_0f_1f_2\ldots$ with $f_t\in\mathbb{D}=\mathcal{F}$ can be addressed. 
In this paper, we take a significant step further by incorporating the Koopman operator $\mathbf{K}$ as an intrinsic component of the proposed temporal logic.  
That is, as with the until operator, we propose to include the Koopman operator $\mathbf{K}$ itself in the alphabet of KTL. 
Aside from including $\mathbf{K}$, the alphabet of KTL is basically the same as for STL. 
%KTL is defined over observables at a single time step. 
%Then, we consider a set of observables $V=\{f^{(1)},\ldots,f^{(p)}\}$. 
%These observables are chosen to extract the spatio properties of interest from the dynamical system. 
%We denote a space spanned by $V$ as $\mathcal{D} = \mathcal{F}\times\ldots\times\mathcal{F}$. 
The grammar of KTL is the same as that of STL, presented as follows:
\begin{definition}[Syntax] \label{def:ktl-syntax}
    The basic formulae of KTL are defined by the grammar
    \begin{align*}
        \varphi \Coloneqq & \mu \mid \neg\,\varphi \mid \varphi_1 \land \varphi_2 \mid \varphi_1 \lor \varphi_2 \mid
        %\mid \mathbf{F}_{[a,b]}\varphi \mid \mathbf{G}_{[a,b]}\varphi \mid 
        \varphi_1 \,\mathbf{U}_{[a,b]}\, \varphi_2.
    \end{align*}
\end{definition}
In KTL, an atomic proposition is interpreted as a predicate, denoted by $\mu$. 
The predicate $\mu:\mathbb{D}=\mathcal{F} \to \mathbb{B}$ is again an interface between an observable-valued signal and a truth value. 
In this paper, to control the time evolution %of a scalar-valued output 
of nonlinear systems (see Sections~\ref{MPC} and \ref{application}), as in \cite{alex2020} we choose an \emph{evaluation functional} $\xi_{\bm{x}}$ at state $\bm{x}$ of the system \eqref{eq:discrete-nonlinear-system} as an element of the dual space of $\mathcal{F}$. \footnote{We note that an additional assumption of $\mathcal{F}$ is necessary, for instance, $\mathcal{F}$ is a reproducing kernel Hilbert space.} 
We will make a remark on the choice of the predicate later. 
In the current choice, for an observable $f$ and initial state $\bm{x}$, we have the output $y_t= f(\bm{x}_t)=(\mathbf{K}^t f)(\bm{x})=\xi_{\bm{x}}(\mathbf{K}^tf)$. 
The truth value of $\mu$ is determined by a sign of the evaluation of $\mathbf{K}^{t} f$, i.e., $\mu := \{ \xi_{\bm{x}}(\mathbf{K}^{t} f) \geq 0 \}$. %, where $\xi_{\bm{x}}(f):\mathcal{F}\to\mathbb{R}$ is a functional that substitutes $\bm{x}$ into $f$. 
A KTL formula is then constructed by combining predicates with the Boolean and temporal operators in the same manner as in STL. 
One significant difference from STL is that it is enough to address an (initial) observable $f$, not the associated signal, because the Koopman operator $\mathbf{K}$ as a representation of dynamics is included in this temporal logic. 
%The notions $\neg$ (negation), $\land$ (conjunction), and $\lor$ (disjunction) are the standard boolean operators. 
%The notion $\mathbf{U}_{[a,b]}$ (Until) is a temporal operator. 
%%The index $[a,b]$ represents the time interval between $a$ and $b$. 
\begin{comment}
A predicate $\mu$ is satisfied at time $k$ if $\pi^\mu(w_k)\ge 0$. 
The negation $\neg\varphi$ is satisfied if $\varphi$ is not satisfied, and the conjunction $\varphi_1\wedge\varphi_2$ is satisfied if both $\varphi_1$ and $\varphi_2$ are satisfied. 
The formula $\varphi_1\mathbf{U}_{[a,b]}\varphi_2$ is satisfied at time $k$ if $\varphi_2$ is satisfied at some time $k'\in[k+a,k+b]$ and $\varphi_1$ is satisfied at all times in $[k,k']$. 
The formula $\mathbf{F}_{[a,b]}\varphi$ is satisfied at time $k$ if $\varphi$ is satisfied at some time in $[k+a,k+b]$, and $\mathbf{G}_{[a,b]}\varphi$ is satisfied at time $k$ if $\varphi$ is satisfied at all times in $[k+a,k+b]$. 
\end{comment}
%Given an initial observable $f_0$, the 
Thus, we can state the satisfaction of a KTL formula \emph{by the nonlinear system \eqref{eq:discrete-nonlinear-system}}. 
Precisely, the satisfaction of a KTL formula $\varphi$ by the nonlinear system \eqref{eq:discrete-nonlinear-system} at time $t$ for a given observable $f$ is denoted by $(f,t)\models\varphi$. 
The following definition provides Boolean semantics for the KTL grammar in Definition~\ref{def:ktl-syntax}. 
\begin{definition}[Boolean semantics] \label{def:ktl-boolean-semantics}
    The satisfaction of a KTL formula by the nonlinear system \eqref{eq:discrete-nonlinear-system} at discrete time $t$ for a given observable $f$ is inductively defined as follows: 
    \begin{equation*}
        \begin{alignedat}{2}
        &(f,t)\models\mu
            &&\Leftrightarrow \xi_{\bm{x}}(\mathbf{K}^{t} f) = (\mathbf{K}^{t} f)(\bm{x}) \geq 0,\\
        &(f,t)\models\neg\varphi
            &&\Leftrightarrow (f,t)\not\models\varphi, \\
        &(f,t)\models\varphi_1\wedge\varphi_2
            &&\Leftrightarrow (f,t)\models\varphi_1\wedge(f,t)\models\varphi_2, \\
        &(f,t) \models \varphi_1 \mathbf{U}_{[a,b]} \varphi_2
            &&\Leftrightarrow \exists t'\in[t+a,t+b], (f,t') \models\varphi_2 \\
        & &&\hphantom{{}\Leftrightarrow{}}\wedge\forall t''\in[t,t'], (f,t'')\models\varphi_1. \\
        % &(f_0,k)\models\mathbf{F}_{[a,b]}\varphi
        %     &&\Leftrightarrow \exists k'\in[k+a,k+b], (f_0,k')\models\varphi, \\
        % &(f_0,k)\models\mathbf{G}_{[a,b]}\varphi
        %     &&\Leftrightarrow \forall k'\in[k+a,k+b], (f_0,k')\models\varphi.
        \end{alignedat}
    \end{equation*}
\end{definition}

The validity of a KTL formula is quantitatively evaluated with a functional on the observable space $\mathcal{F}$, which we call a \emph{robustness functional}. 
Given an (initial) observable $f$, the KTL formula $\varphi$ is satisfied by the nonlinear system \eqref{eq:discrete-nonlinear-system} at time $t$, i.e., $(f,t) \models \varphi$ if and only if the corresponding robustness functional $\rho^\varphi$ is non-negative: $\rho^\varphi(f,t) \geq 0$. 
The quantitative semantics of KTL are constructed recursively in the following manner.
\begin{definition}[Quantitative semantics]  \label{def:ktl-quantitative-semantics}
    The complete robustness semantics for KTL are defined %through the evaluation functional $\xi_{\bm{x}}$ 
    as follows:
    \begin{equation*}
    \begin{aligned}
        \rho^{\mu}(f,t) &= \xi_{\bm{x}}(\mathbf{K}^{t} f) = (\mathbf{K}^{t} f)(\bm{x}), \\
        \rho^{\neg\varphi}(f,t) &= -\rho^{\varphi}(f,t),\\
        \rho^{\varphi_1\land\varphi_2}(f,t)
            &= \min\bigl(\rho^{\varphi_1}(f,t),\rho^{\varphi_2}(f,t)\bigr),\\
        \rho^{\varphi_1\mathbf{U}_{[a,b]}\varphi_2}(f,t)
            &= \max_{t'\in[t+a,t+b]}\min\Bigl(\rho^{\varphi_2}(f,t'),\\
            &\qquad
               \min_{t''\in[t,t']}\rho^{\varphi_1}(f,t'')\Bigr).\\
        % \rho^{\mathbf{F}_{[a,b]}\varphi}(w,k)
        %     &= \max_{k'\in[k+a,k+b]}\rho^{\varphi}(f_0,k'),\\
        % \rho^{\mathbf{G}_{[a,b]}\varphi}(w,k)
        %     &= \min_{k'\in[k+a,k+b]}\rho^{\varphi}(f_0,k').
    \end{aligned}
    \end{equation*}
\end{definition}
The value of the robustness functional is determined with the iterated actions of the Koopman operator $\mathbf{K}$. 
The horizon length for KTL is then introduced as the number of the iterated actions of $\mathbf{K}$ required to determine the value of the robustness functional. 
Its definition is the same as %in 
that for STL. 
\begin{remark}

The ability to freely choose the  observable $f$ reflects the flexibility of modeling with KTL. 
Above, we simply introduced $f$ as the output of the nonlinear system \eqref{eq:discrete-nonlinear-system}. 
Needless to say, we can design $f$ to verify and monitor various properties of the dynamics, for instance, a general inequality constraint of the output as given in Section~\ref{specification-example} and in Figure~\ref{fig:monitoring}. 
This indicates that various predicates are described with properly designed $f$, which is directly beneficial for the MPC formulation (see Sections~\ref{MPC} and \ref{application}). 
In addition to such a time-domain property, as in \cite{time-frequency-logic}, we can define a novel predicate to verify \emph{spectral properties} of the nonlinear system \cite{mezic2005,koopman-introduction-springer}, which is enabled for the first time by including the Koopman operator $\mathbf{K}$. 
This will be pursued in our future work. 
\end{remark}

Here, we present an important result of KTL applicable not only to monitoring but also to control in this paper. 
We show that the time evolution of a robustness functional is represented by the Koopman operator $\mathbf{K}$. 
For a KTL formula $\varphi$ and $t\in\mathbb{T}$, a robustness functional can be regarded as a function of the initial state $\bm{x}$, that is,
\begin{equation}
    \label{eq:robustness-definition}
    \rho^\varphi_t(\bm{x}) := \rho^\varphi(f,t) = G\big((\mathbf{K}^{t}f)(\bm{x}), \ldots, (\mathbf{K}^{t+h^\varphi}f)(\bm{x})\big),%=(\mathbf{K}^tf)(\bm{x}),\qquad \bm{x} \in \mathbb{R}^n.
\end{equation}
where $G$ is a composite function of max and min functions associated with the quantitative semantics of the KTL formula $\varphi$ (see Section~\ref{specification-example} for an example of $G$). 
The function $\rho^\varphi_t(\bm{x})$ represents the robustness function of STL at discrete time $t$ of the trajectory of the nonlinear system \eqref{eq:discrete-nonlinear-system} starting from $\bm{x}$. 
The following theorem shows that the robustness function evolves by the action of the Koopman operator. 
\begin{theorem} \label{thm:1}
For a KTL formula $\varphi$ and $t\in\mathbb{T}$, the time evolution of the robustness function $\rho^\varphi_t$ is represented by the Koopman operator $\mathbf{K}$: for the robustness function $\rho^\varphi_0$ at time $t=0$,
\begin{equation} \label{eq:thm1}
    \rho^\varphi_t = \mathbf{K}^t \rho^\varphi_0.
\end{equation}
\end{theorem}
\begin{proof}
By \eqref{eq:koopman-definition} of the Koopman operator, 
$(\mathbf{K}^t g)(\bm{x}) = g \big(\bm{T}^t(\bm{x})\big)$ holds for an observable $g\in\mathcal{F}$.  
Hence, \eqref{eq:thm1} is equivalent to
$\rho^\varphi_t(\bm{x})=\mathbf{K}^t\rho^\varphi_0(\bm{x})=\rho^\varphi_0\big(\bm{T}^t(\bm{x})\big)$ for $\bm{x}\in\mathbb{R}^n$. 
We show the following shift property:
\begin{equation}\label{eq:shift}
  \rho^\varphi_{t+s}(\bm{x}) = \rho^\varphi_t\big(\bm{T}^s(\bm{x})\big),
  \qquad \forall s,t\in\mathbb{T},\ \forall \bm{x}\in\mathbb{R}^n .
\end{equation}
Setting $t=0$ in \eqref{eq:shift} yields \eqref{eq:thm1}. 
We prove \eqref{eq:shift} by induction on the structure of $\varphi$;
that is, we show that \eqref{eq:shift} holds for every predicate $\mu$, and if \eqref{eq:shift} holds for $\varphi$, $\varphi_1$, and $\varphi_2$, then it also holds for $\neg\varphi$, $\varphi_1\wedge\varphi_2$, and $\varphi_1\,\mathbf{U}_{[a,b]}\,\varphi_2$.

\emph{Predicate.} From \eqref{eq:robustness-definition} of $\rho^\varphi_t$ and the definition of the robustness functional $\rho^\mu$ in Definition~\ref{def:ktl-quantitative-semantics}, we have $\rho^\mu_{t+s}(\bm{x}) = \mathbf{K}^{t+s}f(\bm{x}) = f\big(\bm{T}^{t+s}(\bm{x})\big)$. 
Then, the following relation holds:
\begin{equation*}
    f\big(\bm{T}^{t+s}(\bm{x})\big)
    = f\big(\bm{T}^{t}(\bm{T}^s(\bm{x}))\big)
    = (\mathbf{K}^t f)\big(\bm{T}^s(\bm{x})\big)
    = \rho^\mu_t\big(\bm{T}^s(\bm{x})\big).
\end{equation*}

\emph{Negation.} By Definition~\ref{def:ktl-quantitative-semantics} and the induction hypothesis \eqref{eq:shift}, we have
\begin{equation*}
  \rho^{\neg\varphi}_{t+s}(\bm{x})
  = -\rho^{\varphi}_{t+s}(\bm{x})
  = -\rho^{\varphi}_{t}\big(\bm{T}^s(\bm{x})\big)
  = \rho^{\neg\varphi}_{t}\big(\bm{T}^s(\bm{x})\big).
\end{equation*}

\emph{Conjunction.} Similarly, we have
\begin{align*}
  \rho^{\varphi_1\wedge\varphi_2}_{t+s}(\bm{x})
  &= \min\big(\rho^{\varphi_1}_{t+s}(\bm{x}),\,
              \rho^{\varphi_2}_{t+s}(\bm{x})\big)\\
  &= \min\Big(\rho^{\varphi_1}_{t}\big(\bm{T}^s(\bm{x})\big),\,
              \rho^{\varphi_2}_{t}\big(\bm{T}^s(\bm{x})\big)\Big) \\
   &= \rho^{\varphi_1\wedge\varphi_2}_{t}\big(\bm{T}^s(\bm{x})\big).
\end{align*}

\emph{Until.} Let $\varphi=\varphi_1\,\mathbf{U}_{[a,b]}\,\varphi_2$.
Substituting $t'=\tau'+s$ and $t''=\tau''+s$ in Definition~\ref{def:ktl-quantitative-semantics} and using the induction hypothesis \eqref{eq:shift}, we have
\begin{align*}
  \rho^\varphi_{t+s}(\bm{x})
  &= \max_{t'\in[t+s+a,\,t+s+b]} \min\Big(\rho^{\varphi_2}_{t'}(\bm{x}),
     \min_{t''\in[t+s,\,t']}\rho^{\varphi_1}_{t''}(\bm{x})\Big)\\
  &= \max_{\tau'\in[t+a,\,t+b]} \min\Big(\rho^{\varphi_2}_{\tau'+s}(\bm{x}),
     \min_{\tau''\in[t,\,\tau']}\rho^{\varphi_1}_{\tau''+s}(\bm{x})\Big)\\
  &= \max_{\tau'\in[t+a,\,t+b]} \min\Big(\rho^{\varphi_2}_{\tau'}\big(\bm{T}^s(\bm{x})\big), \\    &\qquad \min_{\tau''\in[t,\,\tau']}\rho^{\varphi_1}_{\tau''}\big(\bm{T}^s(\bm{x})\big)\Big)\\
  &= \rho^\varphi_t\big(\bm{T}^s(\bm{x})\big).
\end{align*}
% The derived operators $\vee$, $\mathbf{F}_{[a,b]}$, and
% $\mathbf{G}_{[a,b]}$ are expressed by $\neg$, $\wedge$, and $\mathbf{U}_{[a,b]}$. 
Therefore, \eqref{eq:shift} holds for every
KTL formula, which completes the proof.
\end{proof}

\begin{comment}
\begin{remark}
Theorem~\ref{thm:1} states that $\rho^\varphi_t$ is itself an observable whose time evolution is governed by the Koopman operator $\mathbf{K}$. 
\end{remark}
\end{comment}
\begin{comment}
The temporal operator $\mathbf{F}_{h}\varphi$ represents the satisfaction of the formula $\varphi$ by at least one state during the time interval $h$. 
The temporal operator $\mathbf{G}_{h}\varphi$ represents the satisfaction of the formula $\varphi$ by every state during the time interval $h$. 
The temporal operator $\mathbf{U}_{h}\varphi$ represents the satisfaction of the formula $\varphi_2$ by at least one state during the time interval $h$ and the satisfaction of the . 
The meanings of negation, disjunction, and conjunction operator are 
\end{comment}

%%%%%%%%%%%%%%%%%%%%%%%%%%%%%%%%%%%%%%%%%%%%%%%%%%%%%%%%%%%%

\subsection{Specification Example} \label{specification-example}
\begin{figure}[t]
    \centering
    \includegraphics[width=0.98\linewidth]{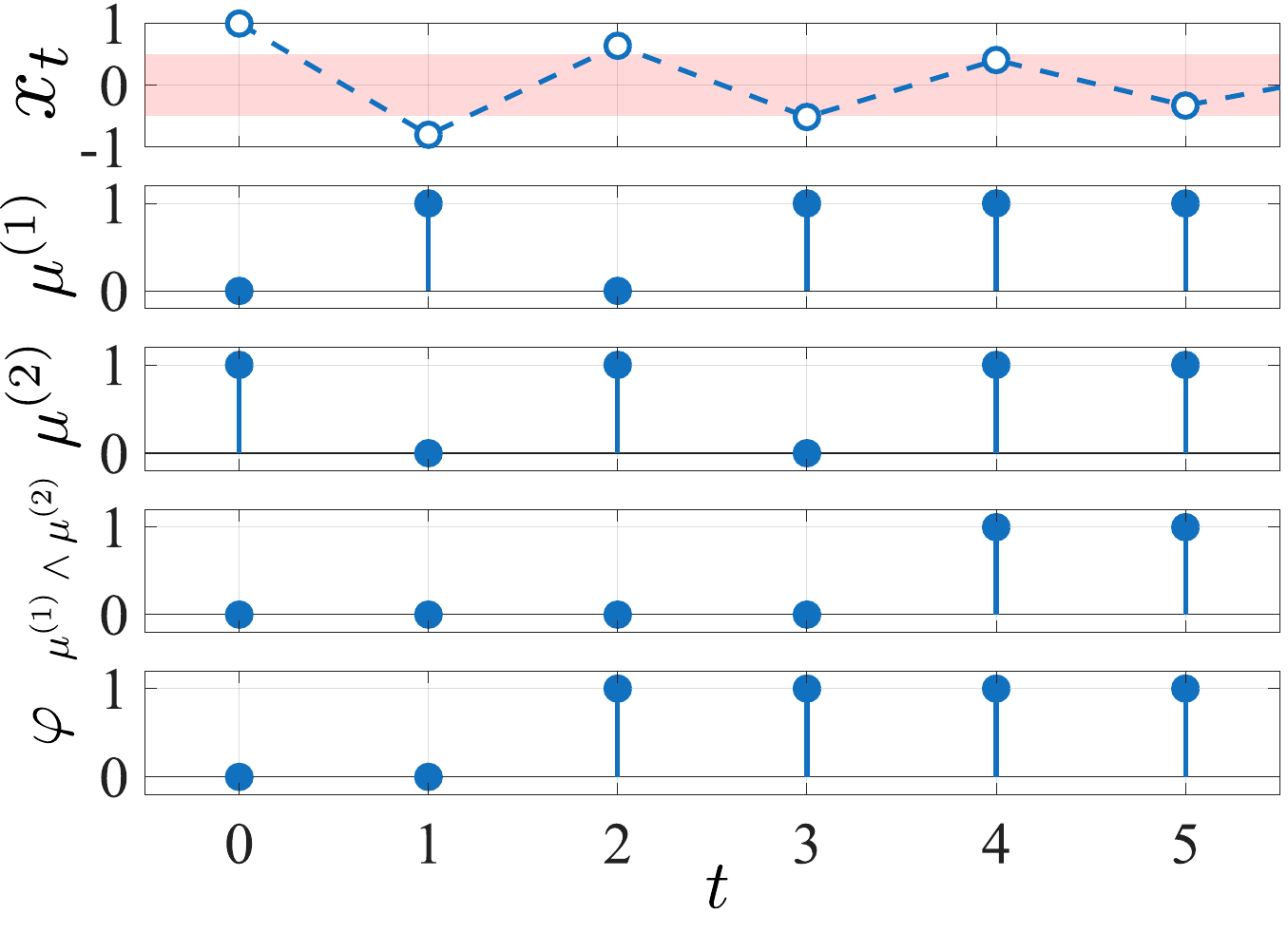}
    \caption{Monitoring result of the dynamical system \eqref{eq:example1-system} against a KTL formula $\varphi=\mathbf{F}_{[0,2]}(\mu^{(1)}\land \mu^{(2)})$, where $\mu^{(1)}=\{\xi_{x_0} (\mathbf{K}^{t} f^{(1)}) \geq0\}$, $\mu^{(2)}=\{\xi_{x_0} (\mathbf{K}^{t} f^{(2)}) \geq0\}$, $f^{(1)}(x)=-x+0.5$ and $f^{(2)}(x)=x+0.5$. A graph in row 1 shows the trajectory of $x_t$, graphs in row 2 and 3 show the truth values of $\mu^{(1)}$ and $\mu^{(2)}$, a graph in row 4 shows the truth value of $\mu^{(1)}\land \mu^{(2)}$, and a graph in row 5 shows the truth value of the KTL formula $\varphi$.}
    \label{fig:monitoring}
\end{figure}

Here, we provide a specification example with KTL. 
Let us consider a simple linear dynamical system, given by
\begin{equation}
    \label{eq:example1-system}
    x_{t+1} = ax_{t}, \quad a=-0.8,
\end{equation}
where $x_{t}\in\mathbb{R}$ represents the one-dimensional state at time $t\in\mathbb{T}$. 
The Koopman operator $\mathbf{K}$ for \eqref{eq:example1-system} is represented as
\begin{equation*}
    (\mathbf{K}^{t} f) (x) = f(a^t x).
\end{equation*}
We here verify the dynamics of \eqref{eq:example1-system} with an initial state $x_{0}=1$ using KTL. 
For this, let us focus on two properties described by the inequality constraints $x\leq0.5$ and $x\geq-0.5$, and introduce the associated observables as $f^{(1)}(x)=-x+0.5$ and $f^{(2)}(x)=x+0.5$. 
The predicates corresponding to these properties are represented as $\mu^{(1)}=\{\xi_{x_0} (\mathbf{K}^{t} f^{(1)}) = (\mathbf{K}^{t} f^{(1)})(x_0)=-a^{t}x_{0}+0.5\geq0\}$ and $\mu^{(2)}=\{\xi_{x_0} (\mathbf{K}^{t} f^{(2)}) = a^{t}x_{0}+0.5\geq0\}$. 
Thus, we aim to verify the temporal property described as the KTL formula $\varphi=\mathbf{F}_{[0,2]}(\mu^{(1)}\land \mu^{(2)})$. 
Note that the corresponding robustness functional is $\rho^\varphi(f,t)=\max_{t'\in[t,t+2]}\min((\mathbf{K}^{t'} f^{(1)})(x_0), (\mathbf{K}^{t'} f^{(2)})(x_0))$. 
This $\varphi$ indicates whether or not the state $x$ is constrained within $|x|\leq0.5$ during three time steps. 
Figure~\ref{fig:monitoring} shows the verification result. 
We can see from the graph in row 4 that the dynamic response generated by the linear system \eqref{eq:example1-system} is constrained within $|x|\leq0.5$ at $t=4$.
The false values of $\varphi$ at $t=0,1$ indicate that the response does not decay into $|x|\leq0.5$ within three steps from $t=0,1$. 
The true values of $\varphi$ at $t=2,3,\ldots$ indicate that the response settles in $|x|\leq0.5$ within three steps from each time step.

%%%%%%%%%%%%%%%%%%%%%%%%%%%%%%%%%%%%%%%%%%%%%%%%%%%%%%%%%%%%%%%%%%%%%%%%%%%%%%%%%%%%%%%%%%%%%%%%%%%%%%%%%%%%%%%%

\section{MODEL PREDICTIVE CONTROL} \label{MPC}
This section presents a formulation of MPC from specifications described by KTL. 

\subsection{Koopman Temporal Logic for Controlled System}
The KTL in Section~\ref{koopman-temporal-logic} was developed with the Koopman operator for the nonlinear system without control input. 
First, we extend it to a controlled system. 
By using the existing theory \cite{mezic2026koopmanoperatortheoryfundamentals}, the Koopman operator can be introduced for the discrete-time controlled system as follows:
\begin{equation}
    \label{eq:discrete-controlled-system}
    \bm{x}_{t+1}=\bm{T}^{\bm{u}_t}(\bm{x}_{t}).
\end{equation}
The symbols $\bm{x}_t\in\mathbb{R}^{n}$ and $\bm{u}_t\in\mathbb{R}^{m}$ represent the state and control input at discrete time $t\in\mathbb{T}$, and $\bm{T}^{\bm{u}_t}:\mathbb{R}^{n}\to\mathbb{R}^{n}$ represents a nonlinear map dependent on the input $\bm{u}_t$. 
As with the uncontrolled system \eqref{eq:discrete-nonlinear-system}, we choose an observable $f:\mathbb{R}^n\to\mathbb{R}$ and analyze the output $y_t = f(\bm{x}_t)$ of the controlled system \eqref{eq:discrete-controlled-system}. 
Then, the time evolution of an observable along \eqref{eq:discrete-controlled-system} is represented by the Koopman operator $\mathbf{K}^{\bm{u}_t}$, defined as 
\begin{equation}
    \label{eq:koopman-controlled-definition}
    (\mathbf{K}^{\bm{u}_{t}} f)(\bm{x}_{t})
    :=f\bigl(\bm{T}^{\bm{u}_t}(\bm{x}_{t})\bigr)
    =f(\bm{x}_{t+1}).
\end{equation}
%Note that the Koopman operator $\mathbf{K}^{\bm{u}_t}$ here is dependent on the input $\bm{u}_t$.

%%%%%%%%%%%%%%%%%%%%%%%%%%%%%%%%%%%%%%%%%%%%%%%%%%%%%%%%

Second, we introduce KTL for the controlled system \eqref{eq:discrete-controlled-system}. 
%Let $\mathbf{u}=\bm{u}_1\bm{u}_2\ldots$ be a sequence of control inputs applied to \eqref{eq:discrete-controlled-system}. 
The syntax of KTL for the controlled system is the same as that of Definition~\ref{def:ktl-syntax} except for the predicate $\mu := \{ \xi_{\bm{x}}(\mathbf{K}^{\bm{u}_{0} }\cdots \mathbf{K}^{\bm{u}_{t-1}} f) \geq 0 \}$. 
The Boolean and quantitative semantics are also the same as those of Definitions~\ref{def:ktl-boolean-semantics} and \ref{def:ktl-quantitative-semantics} except for $(f,t)\models\mu \Leftrightarrow \xi_{\bm{x}}(\mathbf{K}^{\bm{u}_{0}}\cdots\mathbf{K}^{\bm{u}_{t-1}}f) \geq 0$ and $\rho^{\mu}(f,t) = \xi_{\bm{x}}(\mathbf{K}^{\bm{u}_0} \cdots \mathbf{K}^{\bm{u}_{t-1}}f)$. 
%We denote $\mathbf{K}^{\bm{u}_{t-1}}\cdots\mathbf{K}^{\bm{u}_0}$ by $\mathbf{K}^{\bm{u}_{0:t-1}}$ for simplicity. 
%
As with Theorem~\ref{thm:1}, we show that the robustness function evolves as an observable. 
Here, a robustness functional $\rho^\varphi(f,t)$ is represented as $\rho^\varphi(f,t) = G\big((\mathbf{K}^{\bm{u}_{0}}\cdots\mathbf{K}^{\bm{u}_{t-1}}f)(\bm{x}),\ldots, (\mathbf{K}^{\bm{u}_{0}}\cdots\mathbf{K}^{\bm{u}_{t+h^\varphi-1}}f)(\bm{x})\big) = G\big(f(\bm{T}^{\bm{u}_{t-1}} \circ \cdots \circ \bm{T}^{\bm{u}_0}(\bm{x})),\ldots, f(\bm{T}^{\bm{u}_{t+h^\varphi-1}} \circ \cdots \circ \bm{T}^{\bm{u}_0}(\bm{x}))\big)$. 
%the states $\bm{x},\bm{T}^{\bm{u}_{t}}(\bm{x}),\ldots, \bm{T}^{\bm{u}_{t+h^\varphi}}(\bm{x})$, which are dependent on not only the state $\bm{x}$ but also the control inputs $\bm{u}_t,\ldots,\bm{u}_{t+h^\varphi}$. 
Since the maps $\bm{T}^{\bm{u}_0}, \ldots, \bm{T}^{\bm{u}_{t+h^\varphi-1}}$ depend on the control inputs $\bm{u}_{0},\ldots,\bm{u}_{t+h^\varphi-1}$, 
a robustness functional here cannot be regarded as a function of the initial state $\bm{x}$ as in \eqref{eq:robustness-definition}. 
To address this, %we define the extended system whose state is a stack of $\bm{T}^{\bm{u}_{t-1}} \circ \cdots \circ \bm{T}^{\bm{u}_0}(\bm{x}), \ldots, \bm{T}^{\bm{u}_{t+h^\varphi-1}} \circ \cdots \circ \bm{T}^{\bm{u}_0}(\bm{x})$ as
\begin{comment}
Unlike the KTL formulae for the uncontrolled system \eqref{eq:discrete-nonlinear-system} in Section~\ref{koopman-temporal-logic}, the KTL formulae here depend on the control inputs. 
Precisely, the time evolution of the robustness functional from time $t$ to $t+1$ depends on the control inputs $\bm{u}_t,\ldots,\bm{u}_{t+h^\varphi}$ because the robustness functional at time $t$ is composed of the observables between $[t,t+h^\varphi]$, i.e., $f_t,\ldots,f_{t+h^\varphi}$. 
In the following, we denote $h^\varphi$ by $h$ for simplicity. 
To handle the dependence on inputs, 
\end{comment}
we define the extended system of \eqref{eq:discrete-controlled-system}, given by
\begin{equation} 
    \label{eq:extended-system}
    \bar{\bm{x}}_t := [\bm{x}_{t+h}^\top, \ldots, \bm{x}_t^\top]^\top, \qquad \bar{\bm{x}}_{t+1} = \bar{\bm{T}}^{\bm{u}_{t+h}}(\bar{\bm{x}}_t),
\end{equation}
where $\bar{\bm{x}}_t\in\mathbb{R}^{n(h+1)}$ is an extended state at time $t$ and $\bar{\bm{T}}^{\bm{u}}:\mathbb{R}^{n(h+1)}\to\mathbb{R}^{n(h+1)}$ represents a nonlinear map. 
Note that we denote $h^\varphi$ by $h$ for simplicity. 
The nonlinear map is given by $\bar{\bm{T}}^{\bm{u}_{t+h}}:=[\bm{T}^{\bm{u}_{t+h}}(\bm{x}_{t+h})^\top, (\bm{S}_{0}\bar{\bm{x}}_{t})^\top, \ldots,(\bm{S}_{h-1}\bar{\bm{x}}_t)^\top]^\top$, where $\bm{S}_i=[\bm{0}_{n\times i n}, \bm{I}, \bm{0}_{n\times (h-i)n}] ~ (i=0,\ldots,h-1)$ is a matrix. 
Since the extended state $\bar{\bm{x}}$ represents the stack of $\bm{T}^{\bm{u}_{t-1}} \circ \cdots \circ \bm{T}^{\bm{u}_0}(\bm{x}), \ldots, \bm{T}^{\bm{u}_{t+h^\varphi-1}} \circ \cdots \circ \bm{T}^{\bm{u}_0}(\bm{x})$, the robustness functional $\rho^\varphi(f,t)$ can be regarded as a function of the extended state $\bar{\bm{x}}$, that is
%Then, a robustness functional can be regarded as a function of the initial extended state $\bar{\bm{x}}$, that is, 
\begin{align*}
    \bar\rho^\varphi_t(\bar{\bm{x}}) :=  \rho^\varphi(f,t).
\end{align*}
The function $\bar{\rho}^\varphi_t(\bar{\bm{x}})$ represents the robustness function of STL at time $t$ of the trajectory of the extended system \eqref{eq:extended-system} starting from $\bar{\bm{x}}$. 
Let $\bar{\mathbf{K}}^{\bm{u}}$ be the Koopman
operator for the extended system \eqref{eq:extended-system}. 
We set an observable $\bar{f}: \mathbb{R}^{n(h+1)}\to\mathbb{R}$ for the extended system \eqref{eq:extended-system}. 
Furthermore, we introduce the robustness functional $\bar{\rho}^\varphi(\bar{f},t)$ for \eqref{eq:extended-system}. 
The following theorem shows that the robustness function for the controlled system also evolves by the action of the Koopman operators $\bar{\mathbf{K}}^{\bm{u}}$ parametrized by the inputs $\bm{u}$. 
%
\begin{comment}
Then, because of $\bar{\rho}^\varphi(\bar{f},t)=\rho^\varphi(f\circ\bm{S},t)$, we have  $\bar{\rho}^\varphi(\bar{f}_0,t)|_{\xi_{\bar{\bm{x}}}=\xi_{\bar{\bm{x}}_0}}=\rho^\varphi(f_0\circ\bm{S},t)|_{\xi_{\bm{x}}=\xi_{\bm{x}_0}}$. 
As in Section~\ref{koopman-temporal-logic}, the robustness functional $\rho^\varphi(f_0,t)$ is regarded as a function of an initial state $\bar{\bm{x}}_0$ for the extended system as follows:
\begin{equation*}
    \bar\rho^\varphi_t(\bar{\bm{x}}_0) := \rho^\varphi(f_0,t) \big|_{\xi_{\bm{x}}=\xi_{\bm{x}_0}} 
    = \rho^\varphi(\bar{f}_0,t) \big|_{\xi_{\bar{\bm{x}}}=\xi_{\bar{\bm{x}}_0}} .
\end{equation*}
\end{comment}
%
\begin{theorem} \label{thm:2}
For a KTL formula $\varphi$, $t\in\mathbb{T}$, and given input sequence $\bm{u}_{h},\bm{u}_{h+1},\ldots,\bm{u}_{t+h-1}$, the time evolution of the robustness function $\bar\rho^\varphi_t$ is represented by the Koopman operators $\bar{\mathbf{K}}^{\bm{u}_h},\ldots,\bar{\mathbf{K}}^{\bm{u}_{t+h-1}}$: for the robustness function $\bar{\rho}^\varphi_0$ at time $t=0$, 
\begin{equation} \label{eq:thm2}
    \bar\rho^\varphi_t = \bar{\mathbf{K}}^{\bm{u}_h}\cdots\bar{\mathbf{K}}^{\bm{u}_{t+h-1}} \bar{\rho}^\varphi_0.
\end{equation}
\end{theorem}
\begin{proof}
The proof follows that of Theorem~\ref{thm:1} by replacing the system \eqref{eq:discrete-nonlinear-system} with \eqref{eq:extended-system}, $f$ with $\bar{f}$, $\rho^\varphi$ with $\bar{\rho}^\varphi$, and $\mathbf{K}^t$ with
$\bar{\mathbf{K}}^{\bm{u}_h}\cdots\bar{\mathbf{K}}^{\bm{u}_{t+h-1}}$. 
\end{proof}
\begin{remark}
    Theorem~\ref{thm:2} states that the robustness function $\bar{\rho}^\varphi_t$  is an observable whose time evolution is governed by the Koopman operator. 
    This implies that the Koopman operator can predict the quantitative satisfaction values of KTL formulae for the controlled system \eqref{eq:discrete-controlled-system}.
    %In other words, the Koopman operator is a predictor for the KTL formula against the system \eqref{eq:discrete-controlled-system}. 
    %Note that the Koopman operators in Theorem~\ref{thm:2} are for the extended system \eqref{eq:extended-system}, not for the autonomous system \eqref{eq:discrete-nonlinear-system}. 
\end{remark}

\begin{comment}
\begin{remark}
For $\bar{\bm{x}}_0$ constructed from a trajectory of (8), we have
$\rho^\varphi_t = \bar\rho^\varphi_t(\bar{\bm{x}}_0)$; hence the
robustness of (8) is predicted by \eqref{eq:thm2}. Moreover, for
$\tau\le h$, the shift structure of \eqref{eq:ext} yields
$\bar{\mathbf{K}}^{\bm{u}_h}\cdots\bar{\mathbf{K}}^{\bm{u}_{\tau+h-1}}
\bar{f}_0 = f_0\circ\bm{S}_{h-\tau}$, which is independent of the inputs.
Therefore, $\bar\rho^\varphi_0$ is an observable of \eqref{eq:ext}.
\end{remark}
\end{comment}

%%%%%%%%%%%%%%%%%%%%%%%%%%%%%%%%%%%%%%%%%%%%%%%%%%%%%%%%%%%%%%%%%%%%%%%%%%%%%%%%%%%%%%%%%%%%%%%%%%%%%%%%%%%%%%%%

\subsection{Formulation from Koopman Temporal Logic} \label{control-synthesis}

We now present a formulation of MPC from specifications described by KTL. 
The basic idea is to incorporate the Koopman operator, as a predictor of the quantitative satisfaction value of a given KTL formula, into the optimization problem. 
In the {K-MPC} framework, the Koopman operator is approximated to a finite-dimensional matrix and used as a prediction model (see \cite{KORDA2018149} for details). 
Here, we use the bilinear model in \cite{Bruder2021} as an approximate state-space model of the Koopman operator $\bar{\mathbf{K}}^{\bm{u}_{t}}$ for the extended system \eqref{eq:extended-system}, given by
\begin{equation}
    \label{eq:koopman-bilinear}
    \bm{z}_{t+1} = \bm{A}\bm{z}_t + \bm{B}\bm{u}_{t} + \sum_{i=1}^{m} u_{t}^{(i)} \bm{B}^{(i)} \bm{z}_t.
\end{equation}
The new state $\bm{z}_t:=[\bar{f}^{(1)}(\bar{\bm{x}}_{t-h}), \ldots, \bar{f}^{(N)}(\bar{\bm{x}}_{t-h}), \bar\rho^\varphi_0(\bar{\bm{x}}_{t-h})]^\top$ is defined with $N$ observables %$\bar{f}^{(i)}=f^{(i)}\circ\bm{S}$ 
and the robustness function. 
Including the robustness function here, which is called a dictionary or lifting function, is a novel point in this formulation, so that we can generate a control input designed to satisfy the corresponding formula. 
The matrices $\bm{A}\in\mathbb{R}^{N\times N}$, $\bm{B}\in\mathbb{R}^{N\times m}$, and $\bm{B}^{(i)}\in\mathbb{R}^{N\times N} ~ (i=1,\ldots,m)$ are estimated from time-series data of the original system \eqref{eq:discrete-controlled-system}. 
%By obtaining a long trajectory of the original system, 
The MPC controller then solves at each time step $t$ of the closed-loop operation the following optimization problem:
\begin{mini!}|s|[3]
    {\substack{\bm{u}_{0:h_{\rm P}}}}
    {J\left( \bm{z}_{0:h_{\rm P}}, \bm{u}_{0:h_{\rm P}} \right) \label{eq:kmpcktl1-cost}}
    {\label{eq:kmpcktl1}}
    {}
    \addConstraint{\bm{z}_{\ell+1}} {= \bm{A}\bm{z}_\ell + \bm{B}\bm{u}_\ell + \sum_{i=1}^{m} u_\ell^{(i)} \bm{B}^{(i)} \bm{z}_0 \label{eq:kmpcktl1-predictor}}
    \addConstraint{\bm{z}_0} {= [ \bar{f}^{(1)}(\bar{\bm{x}}_{t-h}), \ldots, \bar{f}^{(N)}(\bar{\bm{x}}_{t-h}), \bar{\rho}^\varphi_{0}(\bar{\bm{x}}_{t-h}) ]^\top  \label{eq:kmpcktl1-initial-condition}}
    %\addConstraint{u_{\min}} {\leq u_i \leq u_{\max}  \label{eq:kmpcdtl1-input-const}}
    % \addConstraint{\left.\begin{aligned}  (\bm{X}_k,\ell)&\models \mathbf{N}^{\bm{u}_0}\varphi\\
    % &\vdots \\
    % (\bm{X}_k,\ell)&\models \mathbf{N}^{\bm{u}_{h_{\rm P}}} \mathbf{N}^{\bm{u}_{h_{\rm P}-1}} \cdots \mathbf{N}^{\bm{u}_0}\varphi 
    % \end{aligned}\right\} \label{eq:kmpcdtl1-dtl} }
    \addConstraint{\bm{z}^{(N+1)}_{\ell} \geq 0 \label{eq:kmpcktl1-ktl} }
    \addConstraint{(\ell=0,\ldots,h_{\rm P}) \notag}
\end{mini!}
The symbol $h_{\rm P}$ is the prediction horizon, and $J$ is the cost function. 
At time $t$, the decision variable $\bm{u}_{0}$ is applied to the plant described by \eqref{eq:discrete-controlled-system}. 
The equality constraint \eqref{eq:kmpcktl1-predictor} is a predictive model based on the Koopman operator. 
The approximation \eqref{eq:koopman-bilinear} induces a non-convex optimization problem in the {K-MPC}, hence we fix $\bm{z}_\ell$ of the bilinear term to the initial value $\bm{z}_0$ over the entire prediction horizon, as adopted in \cite{Bruder2021}. 
The initial value $\bm{z}_0$ is set as \eqref{eq:kmpcktl1-initial-condition}. 
%Functions $\bar{f}^{(1)},\ldots,\bar{f}^{(N)}$ are observables for the extended system \eqref{eq:extended-system}. 
%Here, the robustness function $\rho^\varphi$ corresponding to the KTL formula $\varphi$ is incorporated into the Koopman model. 
%The initial value for predicting $\rho^\varphi$ is set as $\rho^\varphi(w,k-h^\varphi)$, which can be calculated from the past states $\bm{x}_{k-h^\varphi:k}$. 
%Then, the KTL satisfaction is predicted by the Koopman model \eqref{eq:kmpcktl1-predictor}. 
%The input constraint is introduced as \eqref{eq:kmpcdtl1-input-const}. 
%Eqs.~\eqref{eq:kmpcdtl1-cost}, \eqref{eq:kmpcdtl1-predictor}, \eqref{eq:kmpcdtl1-initial-condition}, \eqref{eq:kmpcdtl1-input-const} are the same as those for the K-MPC \cite{KORDA2018149}. 
Equation~\eqref{eq:kmpcktl1-ktl} enforces to satisfy the KTL specification through the robustness function.
\begin{remark}
    The optimization problem \eqref{eq:kmpcktl1} includes \emph{no} binary variables, which is an advantage of MPC with KTL specifications compared to MPC with STL specifications \cite{raman2014,Miyashita2022,hirose2026formal}. 
    This potentially speeds up the optimization, as shown in Section~\ref{numerical-result}.
\end{remark}

%%%%%%%%%%%%%%%%%%%%%%%%%%%%%%%%%%%%%%%%%%%%%%%%%%%%%%%%%%%%%%%%%%%%%%%%%%%%%%%%%%%%%%%%%%%%%%%%%%%%%%%%%%%%%%%%%%%%%%%%

\section{DEMONSTRATION} \label{application}

In this section, as an application of MPC with KTL specifications, we describe by KTL a specification for an AC-DC converter and perform a control simulation. 
The plant's setting is mainly based on our previous paper  \cite{hirose2026formal}. 
% the More-Electric Aircraft (MEA) {\color{red}(see, e.g., [REF])}.
% MEAはIntroで言及することにします。

\begin{figure}[t]
    %\vspace{3.7pt}
    \centering
    \includegraphics[width=1.0\linewidth]{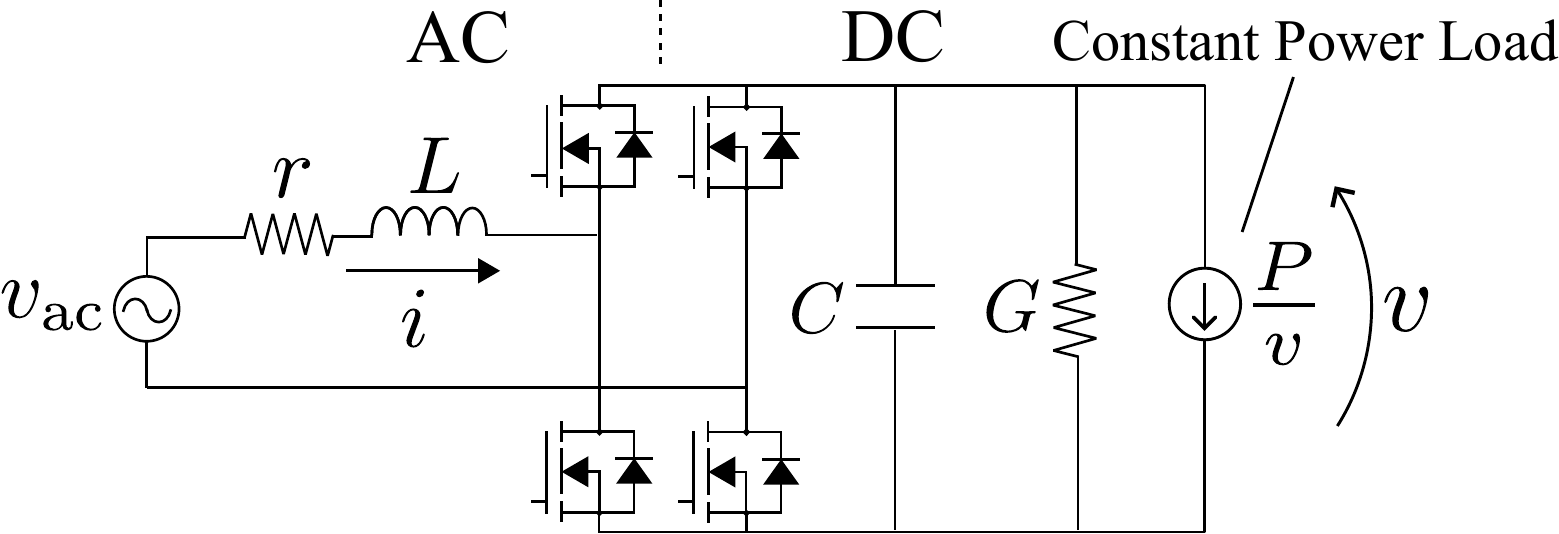}
    \caption{Basic configuration of an AC-DC converter. The Constant Power Load (CPL) is represented by the symbol of a current source.}
    \label{fig:AC-DC-converter}
\end{figure}

\subsection{AC-DC Converter} \label{AC-DC-Converter}

Figure~\ref{fig:AC-DC-converter} shows a single-phase AC-DC converter containing a full-bridge boost rectifier. 
The state variables are defined as $\bm{x}(\tau)=[i(\tau),v(\tau)]^\top$ in continuous time $\tau$, where $i(\tau)$ is the AC current and $v(\tau)$ is the DC voltage. 
The input variable is the duty ratio $s(\tau)\in[-1,1]$. 
The AC supply voltage $v_{\rm ac}(\tau)$ is represented as $v_{\rm ac}(\tau) = E\sin(\omega \tau)$ with the amplitude $E=160/\sqrt{3}\,\rm V$ and the angular frequency $\omega=2\pi\cdot400\,\rm rad/s$ (period $T=2.5\,\rm ms$). 
The DC loads consist of a resistive load of $G=1/47\,\rm S$ and a constant power load (CPL) of $P=1.5\,\rm kW$.  
The current of the CPL is represented as $i_{\rm CPL}(\tau) =P/v(\tau)$. 
The inductance is $L=20\,\mu\rm H$, the capacitance $C=1.2\,\rm mF$, and the resistance $r=0.2\,\Omega$. 
The parameters used in this paper are mainly from \cite{modeling-and-impedance}, which presents an experiment for the MEA. 
The dynamical model of the AC-DC converter is represented by% the nonlinear control-affine system as follows:
\begin{equation}
    \label{eq:AC-DC-converter1}
    \left.
    \begin{alignedat}{2}
        -L\frac{\mathrm{d}i(\tau)}{\mathrm{d}\tau} &= s(\tau)v(\tau) + ri(\tau) - E\sin(\omega \tau) \\
        C\frac{\mathrm{d}v(\tau)}{\mathrm{d}\tau} &= s(\tau)i(\tau) - Gv(\tau) - \frac{P}{v(\tau)} 
    \end{alignedat}
    \right\}.
\end{equation}

The control objectives for the AC-DC converter are listed as follows: 1) The mean value of the DC voltage $v(\tau)$ is equal to the desired value $V_{\rm d}=270\,\rm V$; 2) The power factor of the converter is equal to one; precisely, $i(\tau)$ becomes $I_{\rm d}\sin(\omega \tau)$ with constant amplitude $I_{\rm d}$. 
Here, $I_{\rm d}$ is calculated as $I_{\rm d}=79.9\,\rm A$ from a power-balance equation between the AC source and the DC loads (see \cite{hirose-isie2026} for the detailed calculation).

We impose the following specifications, which will be described with KTL in Section~\ref{KTL-for-converter}, to ensure reliable control of the AC-DC converter. 
First, if the overvoltage occurs, then the short-time average of the DC voltage $v(\tau)$ over the AC period $T$ must return to $280\,\rm V$ within $5\,\rm ms$, namely two cycles of the AC supply voltage. 
This is set in accordance with MIL-STD-704F \cite{MIL-STD-704F} for the MEA. 
Second, the amplitude of the AC current $i(\tau)$ must be maintained above the minimum amplitude $40\,\rm A$. 
Once an overvoltage occurs on the DC side, the AC current $i(\tau)$ decreases to $0\,\rm A$ to reduce the power supplied to the DC loads. 
This sudden stop of the power supply, regarded as load shedding, is harmful to the AC source. 
The minimum amplitude of the AC current $i(\tau)$ prevents sudden load shedding for the AC source.

%%%%%%%%%%%%%%%%%%%%%%%%%%%%%%%%%%%%%%%%%%%%%%%%%%%%%%%%

\subsection{Specifications by Koopman Temporal Logic} \label{KTL-for-converter}

We here describe specifications for the AC-DC converter in Section~\ref{AC-DC-Converter} using KTL.
To this end, we introduce observables for the AC-DC converter. 
The control objectives and specifications are concerned with the amplitude and phase of the AC current $i(\tau)$ and with the mean value of the DC voltage $v(\tau)$, which are not instantaneous values of $i(\tau)$ and $v(\tau)$. 
To address these, we proposed in \cite{hirose-isie2026,hirose2026formal} using a dynamical map, specifically Generalized State-Space Averaging (GSSA), to synthesize observables. 
Note that the choice of observables enables {K-MPC} to perform successfully in an experimental circuit \cite{hirose-isie2026}. 
The index-$h\in\{0,1\}$ GSSA for state $x(\tau)$ is introduced as $\langle x \rangle^{(h)}(\tau) := \int_{\tau-T}^{\tau} x(\tau')\exp({-\mathrm{j}\omega h\tau'}){\rm d}\tau'/T$,
% \begin{equation}
%     \label{eq:GSSA-definition}
%     \langle x \rangle_{h}(t) := \frac{1}{T} \int_{t-T}^{t} x(\tau)\exp({-\mathrm{j}\omega h\tau}){\rm d}\tau,
% \end{equation}
where $T$ is the period of the AC voltage. 
That is, $\langle x \rangle^{(h)}(\tau)$ at time $\tau$ is the moving harmonic-average of the state $x(\tau)$ over $[\tau-T,\tau]$. %during the time between $t-T$ and $t$.
In this paper, we compute GSSA at every AC period $T$ and label $\langle x \rangle^{(h)}(tT)$ as $\langle x \rangle^{(h)}_{t}$ by using the discrete time $t$. 
The $\langle i \rangle^{(1)}$ is related to the phasor of the AC current $i(\tau)$, and $\langle v \rangle^{(0)}$ is the mean value of $v(\tau)$. 
The amplitude of the AC current $i(\tau)$ is represented as $2|\langle i \rangle^{(1)}|$. 
Because the basic properties of converter's specifications are $\langle v \rangle^{(0)} \leq 280\,\rm V$ and $2|\langle i \rangle^{(1)}| \geq 40\,\rm A$, 
we introduce observables for KTL as $f^{(1)}(\bm{x})= 280 - \langle v \rangle^{(0)}$ and $f^{(2)}(\bm{x})= 2|\langle i \rangle^{(1)}| - 40$. 
By using these observables, we describe the KTL formula $\varphi$ corresponding to the converter's specifications as
\begin{equation}
    \varphi = (2|\langle i \rangle^{(1)}_{t}|-40 \geq 0) \ \mathbf{U}_{[0,2]} \ (280 - \langle v \rangle^{(0)}_{t} \geq 0).
\end{equation}

%%%%%%%%%%%%%%%%%%%%%%%%%%%%%%%%%%%%%%%%%%%%%%%%%%%%%%%%%%%%%%%%%%%%%%%%%%%%%%%%

\begin{figure*}
    %\vspace{3.4pt}
    \centering
    \includegraphics[width=0.96\linewidth]{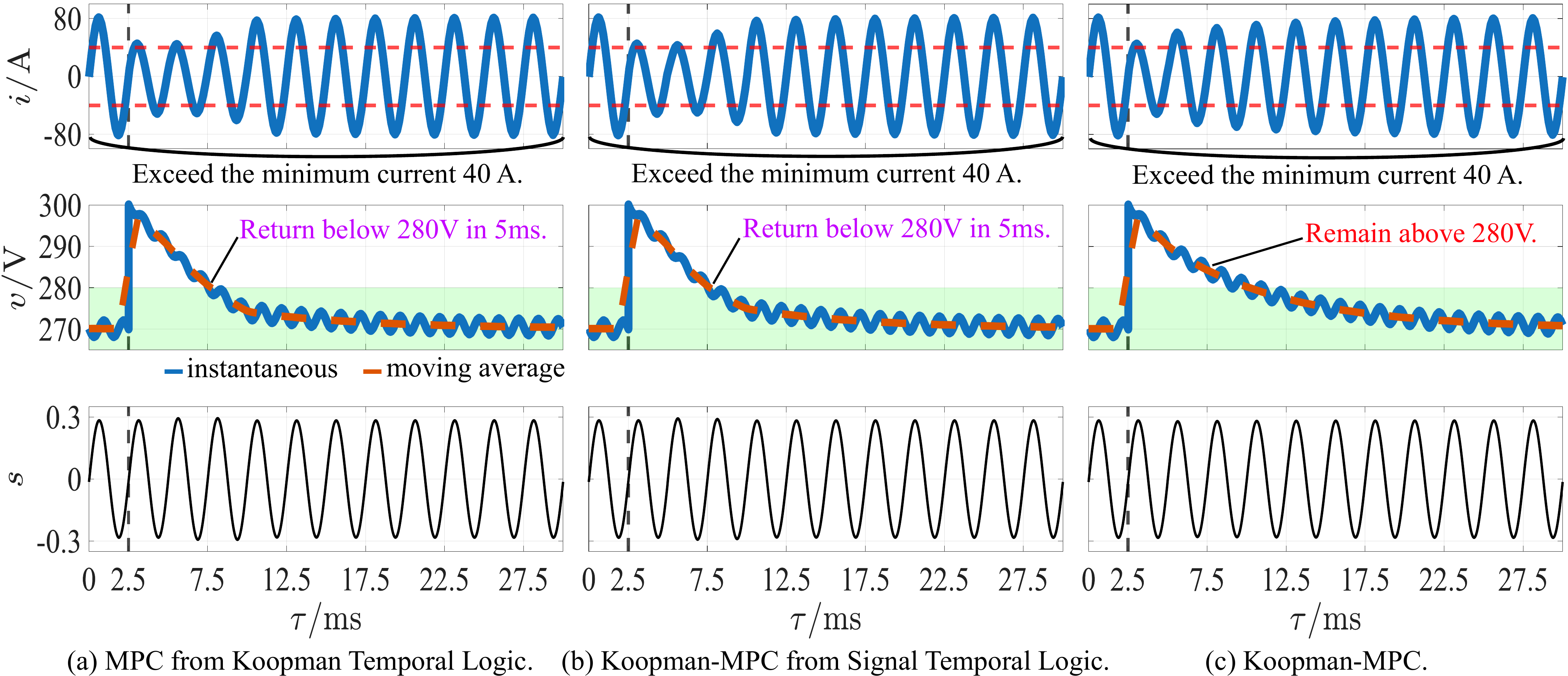}
    \caption{Comparison of the control simulations by MPC from Koopman Temporal Logic specifications, {K-MPC} from Signal Temporal Logic specifications, and {K-MPC}. Graphs in row 1 show the AC current (\emph{blue} line) and the minimum current $\pm40\,\rm A$ (\emph{red dotted} lines), graphs in row 2 show the DC voltage (\emph{blue} line), its moving average (\emph{orange} line), and the range $v\leq280\,\rm V$ (\emph{green} area), and graphs in row 3 the duty ratio. The vertical \emph{dotted} line in each graph represents the moment when the voltage surge occurs. }
    \label{fig:control-results}
\end{figure*}

%%%%%%%%%%%%%%%%%%%%%%%%%%%%%%%%%%%%%%%%%%%%%%%%%%%%%%%%%%%%%%%%%%%%%%%%%%%%%%%%%

\subsection{Simulation Setting}

We explain the simulation setting for MPC. 
First, we derive the Koopman model for the AC-DC converter. 
As mentioned in Section~\ref{KTL-for-converter}, we synthesize observables from GSSA variables.  
%$\bm{z}_t=[\Im\langle i \rangle^{(1)}_t,\Re\langle i \rangle^{(1)}_t,\langle v \rangle^{(0)}_t,\langle v^{-1}\rangle^{(0)}_t, 90 - 2|\langle i \rangle^{(1)}_{t}|, \langle v \rangle^{(0)}_{t} - 250, 90 - 2|\langle i \rangle^{(1)}_{t+1}|, \langle v \rangle^{(0)}_{t+1} - 250, 90 - 2|\langle i \rangle^{(1)}_{t+2}|, \langle v \rangle^{(0)}_{t+2} - 250, \rho^\varphi_t]$
To describe the control objectives, $\bm{z}_t$ includes $\Im\langle i \rangle^{(1)}_t$, $\Re\langle i \rangle^{(1)}_t$, and $\langle v \rangle^{(0)}_t$. 
The reference signals of $i(\tau)$ and $v(\tau)$ are encoded with the GSSA as $\langle I_{\rm d}\sin(\omega t)\rangle_{1}(\tau)=-\mathrm{j}I_{\rm d}/2$ and $\langle V_{\rm d}\rangle_{0}(\tau)=V_{\rm d}$, which become time-invariant and will be used for the MPC. 
To improve the model's accuracy, $\bm{z}_t$ includes $\langle v^{-1}\rangle^{(0)}_t$ motivated by the presence of the nonlinear term $P/v$ in \eqref{eq:AC-DC-converter1}. 
The predicates $2|\langle i \rangle^{(1)}_{t}|-40$ and $280 - \langle v \rangle^{(0)}_{t}$ are also included through the robustness function $\bar{\rho}^\varphi_{0} (\bar{\bm{x}}_{t-2})$ in $\bm{z}_t$. 
%Recalling that the state-space model \eqref{eq:koopman-bilinear} is approximating the Koopman operator for the extended system, we included the above 
Since $\rho^\varphi_{0}(\bar{\bm{x}}_{t-2})$ depends on $2|\langle i \rangle^{(1)}_{t-1}| - 40$,  $280 - \langle v \rangle^{(0)}_{t-1}$,  $2|\langle i \rangle^{(1)}_{t-2}| - 40$, and  $280 - \langle v \rangle^{(0)}_{t-2}$, they are included in $\bm{z}_t$. 
Using the technique in \cite{KORDA2018149}, $\bm{z}_t$  includes $500$ thin plate spline radial basis functions $\|[2|\langle i \rangle^{(1)}_{t}| - 40,280 - \langle v \rangle^{(0)}_{t}]-\bm{c}_{i}\|^2\log \|[2|\langle i \rangle^{(1)}_{t}| - 40,280 - \langle v \rangle^{(0)}_{t}]-\bm{c}_{i}\|~(i=1,\ldots,500)$, where $\bm{c}_i$ is a center selected randomly from a uniform distribution on the unit box. 
Finally, as described in Section~\ref{control-synthesis}, $\bar{\rho}^\varphi_{0} (\bar{\bm{x}}_{t-2})$ is included in $\bm{z}_t$. 
The dimension of $\bm{z}_t$ is $511$, summarized by
\begin{equation*}
\begin{split}
\bm{z}_t = \bigl[\,
  &\Im\langle i\rangle^{(1)}_t,\ \Re\langle i\rangle^{(1)}_t,\
   \langle v\rangle^{(0)}_t,\ \langle v^{-1}\rangle^{(0)}_t,\\
  &2|\langle i\rangle^{(1)}_{t}|-40,\ 280-\langle v\rangle^{(0)}_{t},\\
  &2|\langle i\rangle^{(1)}_{t-1}|-40,\ 280-\langle v\rangle^{(0)}_{t-1},\\
  &2|\langle i\rangle^{(1)}_{t-2}|-40,\ 280-\langle v\rangle^{(0)}_{t-2},\\
  &\text{500 radial basis functions},\
   \bar{\rho}^{\varphi}(\bar{\bm{x}}_{t-2})\,\bigr]^{\top}.
\end{split}
\end{equation*}

Second, we introduce the input variable at discrete time $t$ determined by the MPC controller. 
Following \cite{hirose-isie2026,hirose2026formal}, we represent the duty ratio as a sine wave, given by $s(\tau)=\tilde{u}^{(1)}\sin(\omega \tau) + \tilde{u}^{(2)}\cos(\omega \tau)$, 
where $\tilde{u}^{(1)},\tilde{u}^{(2)}$ are constants at steady states. 
On the basis of \cite{ijcta}, the steady-state values of $\tilde{u}^{(1)},\tilde{u}^{(2)}$ are calculated from the references of $\langle i \rangle^{(1)}$ and $\langle v \rangle^{(0)}$ as follows: $\check{u}^{(1)} = -2(GV_{\rm d}+P/V_{\rm d})/I_{\rm d}$ and $\check{u}^{(2)} = \omega LI_{\rm d}/V_{\rm d}$. 
We then introduce the control variable $\bm{u}=[u^{(1)},u^{(2)}]^\top$ for the bilinear model \eqref{eq:koopman-bilinear} and set $\tilde{u}^{(1)}=\check{u}^{(1)}+u^{(1)}$ and $\tilde{u}^{(2)}=\check{u}^{(2)}+u^{(2)}$. 
In this paper, we regard $u^{(1)},u^{(2)}$ as the control variables that actuate the GSSA variables $\langle i \rangle^{(1)}_t$ and $\langle v \rangle^{(0)}_t$. 
That is, their values at discrete time $t$ (namely, $\bm{u}_t=[u^{(1)}_t,u^{(2)}_t]^\top$) regulate the values of the lifted variable $\bm{z}_{t+1}$ at time $t+1$ as in the bilinear model \eqref{eq:koopman-bilinear}. 
This implies that $s(\tau)$ is updated at every period of the AC voltage. 

Third, to obtain the training datasets, we compute the GSSA variables $\langle i \rangle^{(1)}_t$, $\langle v \rangle^{(0)}_t (t=0,\ldots,13)$  from the state trajectory $\bm{x}(\tau) \, (\tau\in[0,32.5]\,\rm ms)$. 
An initial point $\bm{x}(\tau=0)=[i(\tau=0), v(\tau=0)]^\top$ is generated with uniform random numbers in the range of $i(\tau=0)\in[-50,50]\,\rm A$ and $v(\tau=0)\in[220,320]\,\rm V$. The control variables $\bm{u}_{t=0,\ldots,13}$ are also generated with uniform random numbers in the range of $[-0.01,0.01]$. 
From $\langle i \rangle^{(1)}_t$, $\langle v \rangle^{(0)}_t (t=0,\ldots,13)$, we construct $11$ snapshots of $\{\bm{z}_t,\bm{z}_{t+1},\bm{u}_{t}\}~(t=2,\ldots,12)$ associated with the extended system \eqref{eq:extended-system}. 
By repeating this process $3000$ times, we obtain $33000$ snapshots of $\{\bm{z}_t,\bm{z}_{t+1},\bm{u}_{t}\}$. 
Then, we apply the extended Dynamic Mode Decomposition \cite{edmd-introduction}, thereby identifying the state-space approximation \eqref{eq:koopman-bilinear} of the Koopman operator $\bar{\mathbf{K}}$ (see \cite{Bruder2021} for the detailed description). 

Fourth, we formulate the optimization problem. 
The cost function is set as $J(\bm{z}_{0:h_{\rm P}}, \bm{u}_{0:h_{\rm P}}) = \sum_{\ell=0}^{h_{\rm P}-1} \left\{ (\bm{z}_\ell-\bm{r})^{\top}\bm{Q}(\bm{z}_\ell-\bm{r}) + \varDelta\bm{u}_\ell^{\top} \bm{R} \varDelta\bm{u}_\ell \right\}$, 
where $\bm{r}=[-I_{\rm d}/2, 0, V_{\rm d}, \bm{0}]^\top$ is the reference, $\bm{Q}=\mathrm{diag}[2,2,1,\bm{0}]$ is the state weight matrix, $\bm{R}=\mathrm{diag}[10,10]$ is the input weight matrix, and $\varDelta \bm{u}_{\ell} = \bm{u}_{\ell+1}-\bm{u}_\ell$ is the increment of the input. 
Note that $\bm{0}$ is the $1\times508$ zero vector. 
The cost function aims to penalize the deviation of the GSSA variables $\bm{z}$ from the reference $\bm{r}$ and the increment of the input variables, latter of which is motivated by minimizing the fluctuation of the input time-series in steady state. 
The prediction horizon is set as $h_{\rm P}=5~(12.5\,\rm ms)$ based on MIL-STD-704F \cite{MIL-STD-704F}, where the DC voltage is stipulated to restore in the order of several tens of milliseconds. 
The input $\bm{u}$ is constrained in the range of $[-0.01,0.01]$ in this paper.

Control simulations were conducted with numerical integration of \eqref{eq:AC-DC-converter1}. 
For comparison, we implemented the K-MPC from STL specifications \cite{Miyashita2022} and the standard K-MPC controller as in \cite{hirose2026formal}. 
The optimization problems for the proposed MPC and the K-MPC were solved by using the QP solver, quadprog. 
The optimization problem for the K-MPC from STL specifications was solved by the MIQP solver, Gurobi 13.0.1. 

%%%%%%%%%%%%%%%%%%%%%%%%%%%%%%%%%%%%%%%%%%%%%%%%%%%%%%%%

\subsection{Numerical Results} \label{numerical-result}

Figure~\ref{fig:control-results} shows numerical results of the closed-loop performance of the AC-DC converter. 
A voltage surge is modeled by adding $30\,\rm V$ to the DC voltage $v(\tau)$ at $\tau=2.5\,\rm ms$, denoted by the dotted vertical line in each graph. 
Regarding control objective 1), all of the controllers achieve small steady-state errors for the mean value of $v(\tau)$ in $[0,2.5]\,\rm ms$ before the surge. 
Regarding control objective 2), we compute the power factor during the steady-state $[0,2.5]\,\rm ms$ by using the Hilbert transform. 
The value of power factor is more than $0.99$ for all of the controllers, implying that they achieve the control objectives well before the surge. 
After the voltage surge, all of the controllers restore the DC voltage $v(\tau)$ to the reference value $270\,\rm V$ while maintaining the amplitude of the AC current above $40\,\rm A$. %during the transient response. 
The transient response of the K-MPC differs from those of the proposed MPC and K-MPC from STL specifications. 
For the standard K-MPC in Figure~\ref{fig:control-results}(c), the overvoltage persists for more than $5\,\rm ms$. 
On the other hand, for the proposed MPC using KTL in Figure~\ref{fig:control-results}(a) and the K-MPC from STL specifications in Figure~\ref{fig:control-results}(b), the voltage $v(\tau)$ returns below $280\,\rm V$ within the predefined duration of $5\,\rm ms$. 
This comparison indicates that only the proposed MPC from KTL specifications and K-MPC from STL specifications achieve reliable performance of the AC-DC converter, which is never achieved by the standard K-MPC. 

Finally, we discuss the computational cost of the proposed MPC from KTL specifications, compared to K-MPC from STL specifications. 
The comparison is shown in TABLE~\ref{tab:computational-time}. 
The computational time was measured on a desktop with $2.4\,\rm GHz$ Intel Xeon. 
K-MPC from STL specifications formulates a Mixed-Integer Problem (MIP), which imposes computational burden and can be solved by an MIP solver, Gurobi. 
In contrast, our MPC from KTL specifications and the standard K-MPC formulate a QP (Quadratic Programming) without integer variables, which is computationally efficient. 
Therefore, the average computational time per iteration of the proposed MPC and K-MPC is shorter than that of K-MPC from STL specifications even when using the same solver, Gurobi. 
In addition, the computational time of the proposed MPC is as short as that of the standard K-MPC. 
Consequently, the proposed MPC from KTL specifications achieves control performance comparable to that of K-MPC from STL specifications while retaining the computational efficiency of standard K-MPC.

\begin{table}[t]
    \centering
    \caption{Average computational time per iteration.}
    \label{tab:computational-time}
    \begin{tabular}{lcc}
        \toprule
        Solver & quadprog & Gurobi\\
        \midrule
        \makecell[l]{MPC from Koopman Temporal Logic} & $14.3\,\rm ms$ & $37.4\,\rm ms$ \\
        \addlinespace
        \makecell[l]{{K-MPC} from Signal Temporal Logic} & --- & $77.0\,\rm ms$ \\
        \addlinespace
        {K-MPC} & $11.3\,\rm ms$ & $36.8\,\rm ms$ \\
        \bottomrule
    \end{tabular}
\end{table}

%%%%%%%%%%%%%%%%%%%%%%%%%%%%%%%%%%%%%%%%%%%%%%%%%%%%%%%%%%%%%%%%%%%%%%%%%%%%%%%%%%%%%%%%%%%%%%%%%%%%%%%%%%%%%%%%%%%%%%%%

\section{CONCLUSIONS}

This paper proposed KTL (Koopman Temporal Logic) that combined STL with the Koopman operator. 
We defined the syntax, Boolean semantics, and quantitative semantics for KTL and showed that the Koopman operator can predict the value of quantitative satisfaction of a KTL formula. 
Furthermore, we provided a formulation of MPC from KTL specifications. 
The optimization problem, including the Koopman operator as the predictor for satisfying a KTL formula, is at most a QP without any integer variables.
We applied the proposed MPC formulation to the AC-DC converter and demonstrated both its effectiveness relative to standard K-MPC and its computational efficiency relative to K-MPC from STL specifications. 
Future work includes monitoring nonlinear systems using KTL and an experimental realization of the MPC from KTL specifications for the AC-DC converter.

\addtolength{\textheight}{-12cm}   % This command serves to balance the column lengths

%%%%%%%%%%%%%%%%%%%%%%%%%%%%%%%%%%%%%%%%%%%%%%%%%%%%%%%%%%%%%%%%%%%%%%%%%%%%%%%%

\section*{ACKNOWLEDGMENT}
% 生成AIを使っていれば明記する。少なくとも，第3著者はいくつかの英文の妥当性をチェックするのに使っています。その程度なら不要？

% ＞使用したので明記します。

% チェックでの使用は明記する義務はないようです。

% The use of AI systems for editing and grammar enhancement is common practice and, as such, is generally outside the intent of the above policy. In this case, disclosure as noted above is not required, but recommended.

The authors used Claude Opus 5.5 to enhance the clarity of Theorems~\ref{thm:1} and \ref{thm:2}. 
All contents were reviewed and edited by the authors, who take full responsibility for the final work.

%%%%%%%%%%%%%%%%%%%%%%%%%%%%%%%%%%%%%%%%%%%%%%%%%%%%%%%%%%%%%%%%%%%%%%%%%%%%%%%%

\raggedbottom

%\raggedbottom
% \bibliographystyle{IEEEtran}
% \bibliography{Sankou}

\end{document}